
\documentclass[modern]{aastex631}
\usepackage{comment}
\usepackage{amsmath}
\usepackage{centernot,cancel} 

\newcommand{\pdf}{\mathrm{Pr}}
\newcommand{\FPP}{\mathrm{FPP}}
\newcommand{\TPP}{\mathrm{TPP}}
\newcommand{\FNP}{\mathrm{FNP}}

\newcommand{\CPP}{\mathrm{CPP}}
\newcommand{\Ntot}{N_{\mathrm{tot}}}
\newcommand{\Npos}{N_{\mathrm{pos}}}
\newcommand{\Hnull}{\mathcal{H}_{\mathrm{null}}}
\newcommand{\Hpos}{\mathcal{H}_{\mathrm{aff}}}
\newcommand{\Kpos}{K_{\mathrm{aff}}}
\newcommand{\Knull}{K_{\mathrm{null}}}
\newcommand{\Haliens}{\mathcal{H}_{\mathrm{aliens}}}

\newcommand{\olsi}[1]{\,\overline{\!{#1}}} 

\newcommand{\natur}{\mathcal{H}_{\mathrm{comb.}}}
\newcommand{\unnatur}{\overline{\mathcal{H}_{\mathrm{comb.}}}}

\newcommand{\naturX}{\mathcal{H}_{\mathrm{kn.comb.}}}
\newcommand{\unnaturX}{\overline{\mathcal{H}_{\mathrm{kn.comb.}}}}

\newcommand{\IKNPn}{\mathcal{H}_{\mathrm{IKCP-N}}}
\newcommand{\IKNPa}{\mathcal{H}_{\mathrm{IKCP-A}}}

\definecolor{quotecol}{rgb}{0.1,0.28,0.46}

\shorttitle{Deconstructing Alien Hunting}
\shortauthors{Kipping \& Wright}
\graphicspath{{./}{figures/}}

\begin{document}

\title{Deconstructing Alien Hunting}

\author[0000-0002-4365-7366]{David Kipping}
\affiliation{Department of Astronomy, Columbia University, 550 W 120th Street, New York, NY 10027, USA}

\author[0000-0001-6160-5888]{Jason Wright}
\affiliation{Department of Astronomy \& Astrophysics, The Pennsylvania State University, University Park, State College, PA 16802, USA}



\begin{abstract}
The search for extraterrestrial (alien) life is one of the greatest scientific
quests yet raises fundamental questions about just what we should be looking
for and how. We approach alien hunting from the perspective of an experimenter
engaging in binary classification with some true and confounding positive
probability ($\TPP$ and $\CPP$). We derive the Bayes factor in such a framework
between two competing hypotheses, which we use to classify experiments as either
impotent, imperfect or ideal. Similarly, the experimenter can be classified as
dogmatic, biased or agnostic. We show how the unbounded explanatory and evasion
capability of aliens poses fundamental problems to experiments directly seeking
aliens. Instead, we advocate framing the experiments as looking for that outside
of known processes, which means the hypotheses we test do not directly
concern aliens per se. To connect back to aliens requires a second level of
model selection, for which we derive the final odds ratio in a Bayesian framework.
This reveals that it is fundamentally impossible to ever establish alien life
at some threshold odds ratio, $\mathcal{O}_{\mathrm{crit}}$, unless we deem the
prior probability that some as-yet-undiscovered natural process could explain
the event is less than $(1+\mathcal{O}_{\mathrm{crit}})^{-1}$. This elucidates
how alien hunters need to carefully consider the challenging problem of how
probable unknown unknowns are, such as new physics or chemistry, and how it
is arguably most fruitful to focus on experiments for which our domain knowledge
is thought to be asymptotically complete.
\end{abstract}



\section{Introduction}

Since the dawn of modern astronomy, our efforts to understand the cosmos have
been coupled to questions regarding the existence of life beyond Earth. Within
the scientific literature, there have been multiple past claims for such life
or evidence thereof - although of course none have withstood broader scrutiny.
Some disproven claims include inhabited Moon \citep{herschel:1780}\footnote{Not
to be confused with \textit{The Great Moon Hoax} \citep{moonhoax}.}, and
Martian canals \citep{lowell:1906}, and some disputed possible evidence of alien life
include the Allan Hills 84001 meteorite \citep{mckay:1996}, Venusian phosphine
\citep{greaves:2021} and the unusual properties of ‘Oumuamua \citep{loeb:2021}.

It is not unusual for claims of extraterrestial life to be accompanied by
proclamations of extreme statistical confidence and considerable media fanfare.
For example, in an article describing observations of lunar topology,
\citet{herschel:1780} described ``the great probability, not to say almost
absolute certainty, of her being inhabited''. Similarly, \citet{lowell:1906}
wrote, ``that Mars is inhabited by beings of some sort or other we may consider
as certain as it is uncertain what these beings may be''. In a more recent
example, in the press briefing announcing the discovery of Gliese 581g
\citep{vogt:2010}, the lead author stated\footnote{See \href{https://www.space.com/9225-odds-life-newfound-earth-size-planet-100-percent-astronomer.html}{this link}}
``that the chances of life on this planet are 100 percent''. Manifestly, the
search for life in the cosmos has often been accompanied with a certain
degree of uncharacteristic ``excessive euphoria'' \citep{schenkel:2006} by
traditional scientific standards. This compromise of rigorous objectivity has
arguably contributed to the entire scientific enterprise being placed in the
cross-hairs for cancellation and open criticism \citep{basalla:2007,
cirkovic:2013}, with a prominent example being the termination of NASA's SETI
program by the US Congress in 1993 \citep{garber:1999}.

In order for the programatic search for life to survive the rollercoaster of
economical resources \citep{garber:1999} (as well as public interest;
\citealt{billings:2012}), it's essential that the community apply the same
strict scientific standards as with any other field. Certainly efforts
have already been undertaken in this direction, with \citet{haqq:2012}
exploring the uncertainty in detecting alien artifacts, \citet{catling:2018}
outlines a Bayesian framework for biosignatures and most recently
\citet{lingam:2023} considers the Bayesian evaluation of technosignatures.
Here, we seek a totally general framework for the problem of looking for
life, applicable to anything from UFOs to fossils on Mars, from exoplanet
biosignatures to alien relics. Such a goal not only
advances the scientific enterprise itself, but the trust the public harbour for
its findings. Accordingly, the scientific value of a given experiment within
this domain needs to have well-defined outcomes and benefits. This speaks to
the need for careful appraisal of the corresponding experimental design, where
critics have previously voiced concern with the potential lack of Popperian
falsifiability \citep{editorial:2009}.

To explore the scientific merit of experiments seeking life, we first present a
generalised Bayesian framework for assessing the odds ratio between two
competing hypotheses, conditioned upon some experimental data. Through a
consideration of the specific challenges facing alien hunters, we show how
differing (yet still generalised) experimental setups affect the scientific
usefulness for various cases. Implications are explored towards the end.

\section{Bayesian Experimental Design}

\subsection{Qualifying the Experiment}

Consider an experiment designed to search for the existence of a specific
phenomenon, for a particular target, given some experimental data. For example,
this could be an experiment searching for the existence of an exoplanet (the
phenomenon) in a set of astronomical observations (the data) for a specific
star (the target). After studying the data/observations/measurements,
$\mathcal{D}$, the experimenter proceeds to classify each target as either a
positive detection of the phenomenon, $\mathcal{S}$, else a null detection,
$\bar{\mathcal{S}}$. These are the only two classes possible, and thus the
experimenter is engaged in binary classification, conditional upon some 
data\footnote{Of course, this act of classification will, in general, be imperfect and thus is handled later in our discussion of confounding positive and true positive probabilities.}.
Note that $\bar{\mathcal{S}}$ is defined as the antithesis of $\mathcal{S}$;
it's not a statement of anything more than the fact the phenomenon of
interest was not detected.

Amongst a sample of $\Ntot$ targets, let us denote that a fraction $f$ truly
manifest the phenomenon in question. We assume here that $f$ is a scalar,
rather than an integer (or really a Boolean flag). In other words, our
hypothesis considers that the phenomenon of interest can \textit{sometimes}
operate; it's not compelled to always operate or never operate (although that
it is indeed a possibility).

\begin{figure*}
\begin{center}
\includegraphics[width=15.0cm,angle=0,clip=true]{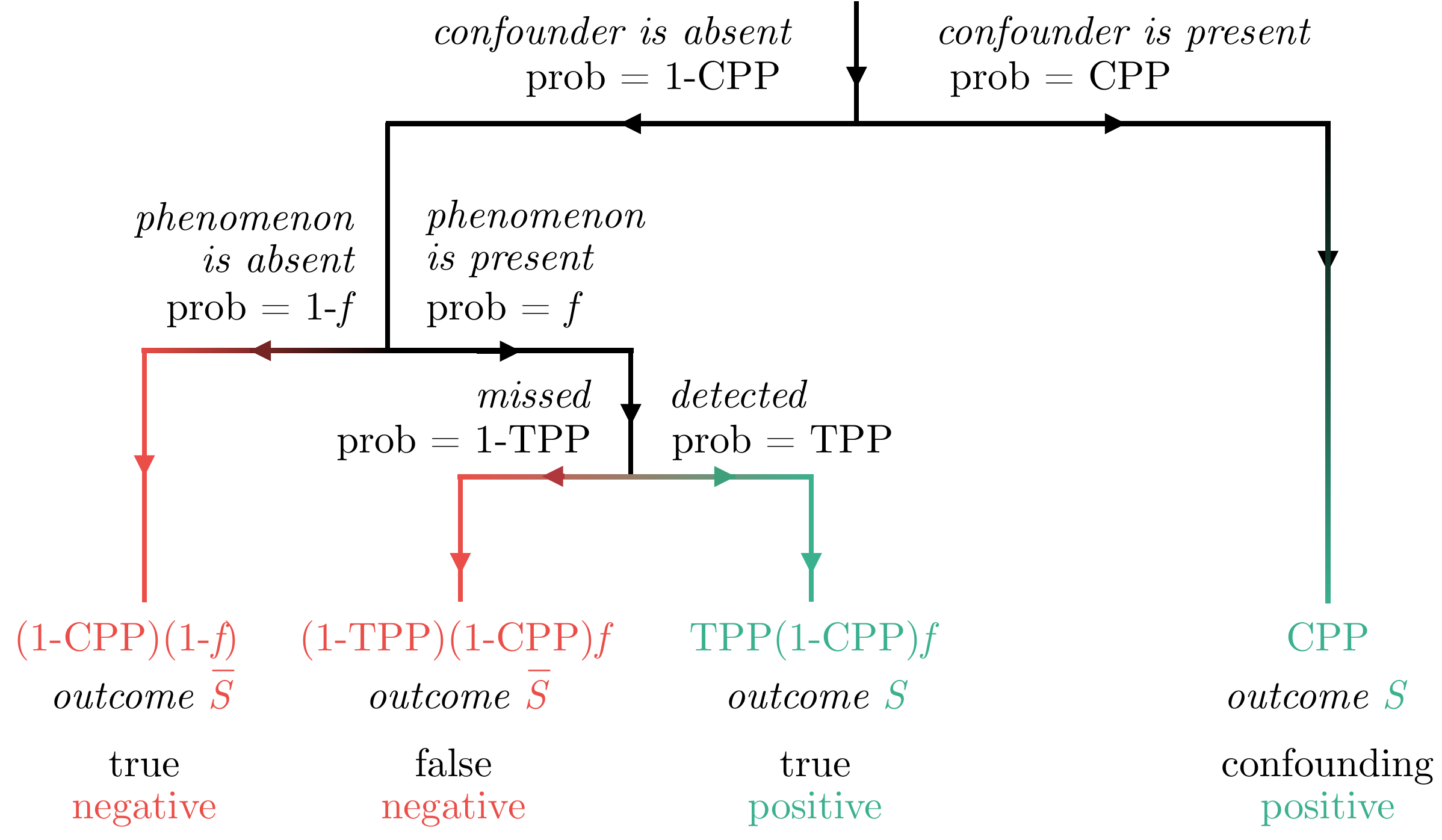}
\caption{
Schematic of the four possible branches resulting from an experiment of a
randomly selected target, where the experiment seeks to detect a phenomenon
which occurs within a fraction $f$ of the targets under consideration,
and the experiment has a confounding positive probability of $\CPP$ and a
true positive probability of $\TPP$.
}
\label{fig:cartoon}
\end{center}
\end{figure*}

A positive detection for a particular target can occur in two distinct ways,
as shown in Figure~\ref{fig:cartoon}. The first is simply that the phenomenon
of interest is genuinely present \textit{and} the experimenter successfully
detects it. The Boolean construction of that statement reveals that the
probability of this occurring must be the product of $f$ with the true positive
probability ($\TPP$); also commonly known as completeness or sensitivity.

It's of course also possible to spuriously obtain a positive detection. We
could define a false positive as the case where we obtain a positive even
though the phenomenon is absent (which has a probability of $1-f$), which
occurs with a frequency given by the false positive probability, $\FPP$. This
means that false positives can only ever happen when the phenomenon is not
present, and thus if $f=1$ it would be impossible to have a false positive. In
practice, \textit{confounding} positives can still occur even when the
phenomenon is present. Whilst this is not technically a false positive, it's
certainly still a spurious result. We thus define the $\CPP$, the confounding
positive rate, to be some background underlying rate at which any experiment
returns a positive. In total, then, the probability of obtaining a positive
detection for a particular target will equal

\begin{align}
\pdf(\mathcal{S}|f) &= \underbrace{ \CPP }_{\mathrm{confounding\,\,positives}} + \underbrace{ (1-\CPP) f \times \TPP }_{\mathrm{true\,\,positives}},
\end{align}

where the $(1-\CPP)$ term ensures $\pdf(\mathcal{S}|f) \in [0,1]$.

\subsection{Likelihood}

Since the labelling by the experimenter is binary, being either $\mathcal{S}$
or $\bar{\mathcal{S}}$, then we have a Bernoulli process where the success
probability equals $\pdf(\mathcal{S}|f)$. Let's make the simplifying assumption
that the $\TPP$ and $\CPP$ are constant for all targets in a given sample, for
which a fraction $f$ genuinely manifest the phenomenon of interest. For a
series of experiments consisting of $\Ntot$ unique and independent targets,
the number of resulting positive detections from these independent Bernoulli
trials must now follow a Binomial distribution. Exploiting this fact, we can
express that the likelihood of obtaining $\Npos$ positive detections from
$\Ntot$ targets, given that a fraction $f$ of the targets manifest the
phenomenon, equals

\begin{align}
\pdf(\Npos|f) =& \binom{\Ntot}{\Npos} \Big( \CPP +  \TPP (1-\CPP) f \Big)^{\Npos}
\Big( 1 - \CPP - \TPP (1-\CPP)f \Big)^{\Ntot-\Npos}.
\label{eqn:likelihood}
\end{align}

\begin{figure*}
\gridline{\fig{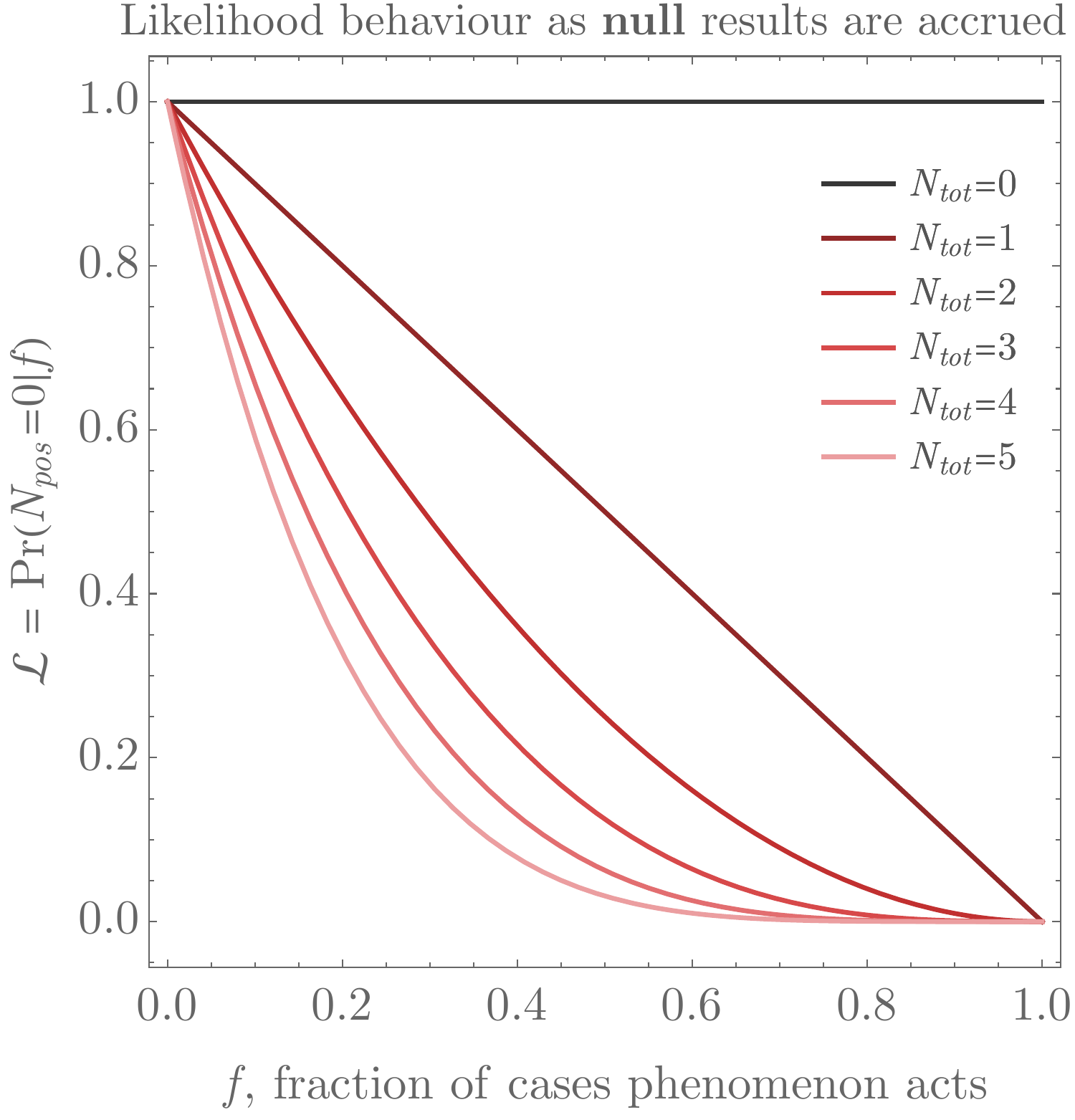}{0.45\textwidth}{(a)}
          \fig{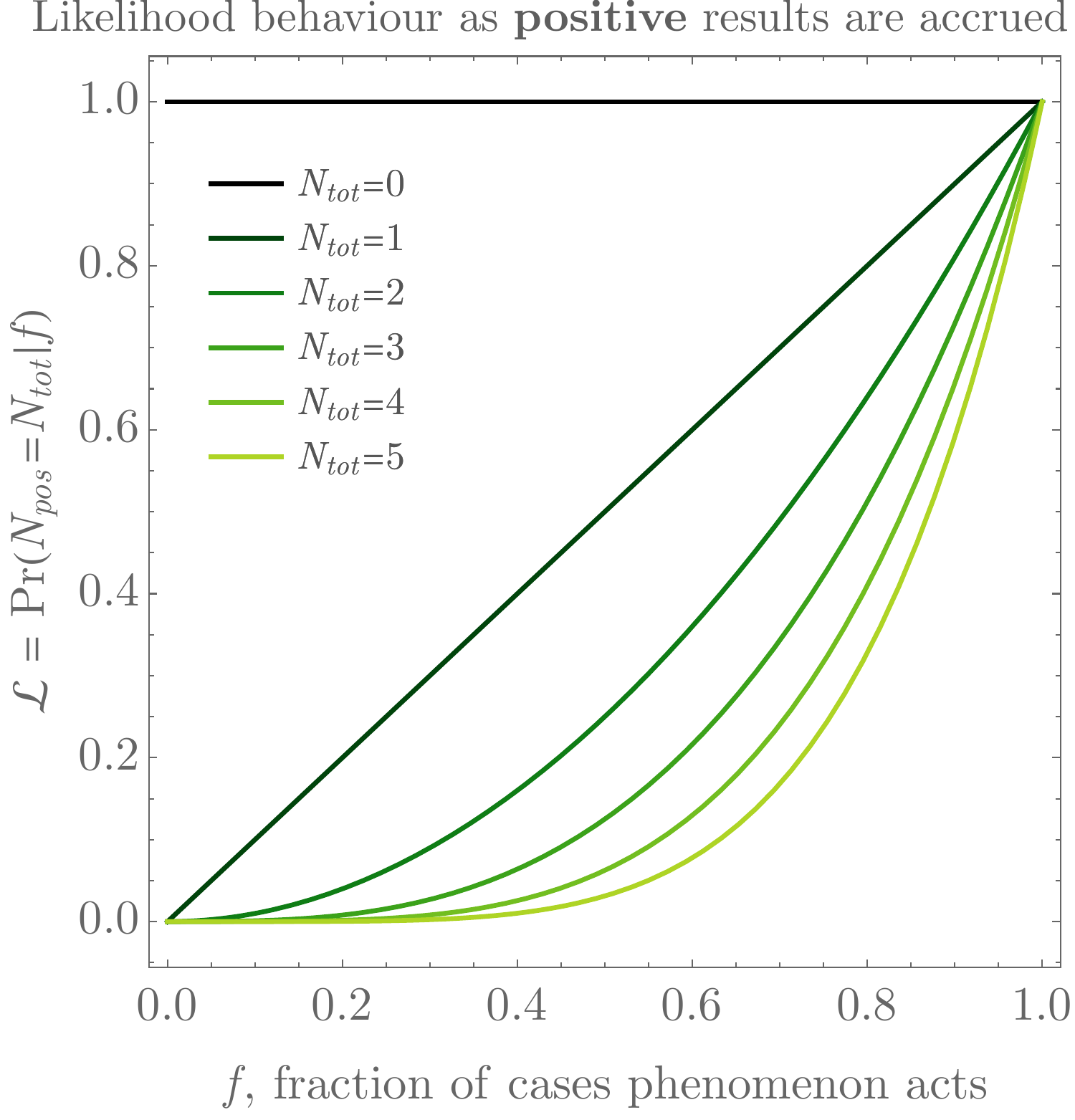}{0.45\textwidth}{(b)}
          }
\caption{
Examples showing the behaviour of the likelihood function, $\pdf(\Npos|f)$
(Equation~\ref{eqn:likelihood}), assuming our data is a series of (a) null
detections, or (b) positive detections. We here assume an
ideal experiment where $\CPP = 0$ and $\TPP = 1$.
\label{fig:likeexamples}}
\end{figure*}

An ideal experiment has a low $\CPP$ and a high $\TPP$, and thus setting 
$\CPP = 0$ and $\TPP = 1$ and inspecting the likelihood behaviour of
Equation~\ref{eqn:likelihood} provides some instructive guidance, as shown in
Figure~\ref{fig:likeexamples}. As expected, a series of null detections (left
panel) leads to a likelihood function peaking at $f=0$, as this is the most
consistent solution with such data. The greater the number of null detections,
the more strongly peaked the likelihood function becomes. In the opposite case
of a series of positive detections (right panel), the maximum likelihood shifts
to $f=1$, again matching our expectations. None of this is new or surprising,
but it establishes how this mathematical formalism has an intuitive
interpretation.

\subsection{Posterior}

A likelihood function describes the probability of obtaining the data, given
some underlying model/hypothesis. So in our case, the likelihood function
describes the probability of obtaining $\Npos$ positive detections, given some
input value for the underlying fraction of targets that manifest the phenomenon
of interest, $f$. Often, the experimenter is more interested in inferring the
actual value of $f$ directly, which requires flipping the conditionals around
via Bayes' theorem:

\begin{align}
\pdf(f|\Npos) =& \frac{\pdf(\Npos|f) \pdf(f)}{\pdf(\Npos)}.
\label{eqn:bayestheorem}
\end{align}

The denominator, $\pdf(\Npos)$, does not depend on $f$ and can be
thought of as a normalizing constant to ensure the probabilities sum to unity:

\begin{align}
\pdf(f|\Npos) =& \frac{\pdf(\Npos|f) \pdf(f)}{ \int_{f=0}^1 \pdf(\Npos|f) \pdf(f) \,\mathrm{d}f }.
\label{eqn:bayestheorem2}
\end{align}

To make progress, we need to adopt a prior for $f$. Ideally, this prior should
be chosen in such a way that it minimally affects the results; i.e. we wish to
avoid imprinting any strong preconceptions about what value $f$ should take.
One obvious choice is a uniform distribution in $f$ between 0 and 1, which
states that all values of $f$ are \textit{a-priori} just as probable as
any other. One might also consider using the so-called Jeffrey's prior
for a Bernoulli process (since $f$ describes a binary outcome), which equates
to $\pdf(f) = 1/(\pi \sqrt{f (1-f)})$ \citep{jeffreys:1946}. Whilst the latter
has a more rigorous justification, we will proceed in what follows with the
uniform prior, since this enables a closed-form solution in what follows
(unlike the Jeffrey's prior). 
With the uniform prior, one finds that the posterior distribution for $f$ is

\begin{align}
\pdf(f|\Npos) =& \frac{
\TPP (\CPP+f \TPP (1-\CPP))^{\Npos} (1-\CPP)^{\Ntot+1-\Npos} (1-f \TPP)^{\Ntot-\Npos}
}{\mathrm{B}_{1-\CPP}[\Ntot+1-\Npos,\Npos+1] - \mathrm{B}_{(1-\TPP)(1-\CPP)}[\Ntot+1-\Npos,\Npos+1]},
\label{eqn:posterior}
\end{align}

where $\mathrm{B}_z[a,b]$ is the incomplete Beta function.

\subsection{Bayesian Model Comparison}

Equipped with a posterior for $f$, we can now proceed to consider questions
fundamental to the scientific endeavour - evaluating hypotheses/models. In
a Bayesian framework, this proceeds through Bayesian model comparison which
compares the relative probability of a particular hypothesis, given the
data, versus the sum of all other competing hypotheses, given the data.
Accordingly, we seek to evaluate

\begin{align}
\frac{\pdf(\mathcal{H}_i|\mathcal{D})}{\sum_{i\ne j}^M \pdf(\mathcal{H}_j|\mathcal{D})},
\end{align}

where we use the symbol $\mathcal{H}$ to denote a hypothesis, and $M$ is
the total number of hypotheses in play. Note that 
$\sum_{i=1}^M \pdf(\mathcal{H}_j|\mathcal{D}) = 1$, since $M$ forms a complete
list of all possible explanations. In what follows, we consider that there are
only two hypotheses in play, $\Hpos$ and $\Hnull$. We define $\Hpos$ as
the ``affirmative'' hypothesis - that the phenomenon in question
\textit{sometimes} acts, such that $f>0$ (but of course must also be $f\leq1$).
Hence, establishing $\Hpos$ would mean that the phenomenon in question does
indeed exist, to some degree.

The null hypothesis, $\Hnull$, can be thought of as a container for the
sum of all alternatives (i.e.
$\sum_{i \ne j}^M \pdf(\mathcal{H}_j|\mathcal{D}))$. But most rigorously,
in order to satisfy that $\sum_{i=1}^M \pdf(\mathcal{H}_j|\mathcal{D}) = 1$,
which here becomes $\pdf(\Hpos|\mathcal{D}) + \pdf(\Hnull|\mathcal{D}) = 1$,
it truly represents the antithesis of $\Hpos$, i.e. $\Hnull\equiv\bar{\Hpos}$.
Connecting this back to our definition that $\Hpos$ represents $0<f\leq1$,
$\Hnull$ is now rigorously defined as $f=0$ - the phenomenon in question never
acts.

A final subtlety is - to what extent are these statements generalisable? Does
adopting $\Hnull$ mean that we truly believe the phenomenon \textit{never}
acts (i.e. anywhere/anytime) or merely that it never acts within
the sample of targets studied? If the sample is considered to be truly
representative of the entire population of potential samples, then indeed
one could plausibly make such a generalisation. However, in practice,
such a strong claim of completely fair representation may not be justifiable.

\subsection{Bayes Factor}

Having defined our hypotheses, we can now proceed with the task of Bayesian
model comparison. The odds ratio of these two hypotheses, given the data, can
be expressed using Bayes' theorem as

\begin{align}
\frac{\pdf(\Hnull|\mathcal{D})}{\pdf(\Hpos|\mathcal{D})} &=
\underbrace{\frac{ \pdf(\mathcal{D}|\Hnull) }{ \pdf(\mathcal{D}|\Hpos) }}_{\mathrm{``Bayes\,\,factor''}}
\overbrace{\frac{ \pdf(\Hnull) }{ \pdf(\Hpos) }}^{\mathrm{\textit{a-priori}\,\,odds\,\,ratio}}
\label{eqn:oddsratio}
\end{align}

In the above, we highlight the ``Bayes factor'', which we will show is directly
calculable from our posterior distribution in Equation~(\ref{eqn:posterior}).
The last term in the above represents our \textit{a-priori} probability
ratio between the two hypotheses. In comparing the Bayes factor and the prior
ratio, it's clear that one depends on the data and the other does not.
Accordingly, only the Bayes factor speaks to what the data, and thus the
experiment, have actually taught us and thus deserves particular attention.

To evaluate the Bayes factor, we exploit the nested nature of our hypotheses
and use the Savage-Dickey theorem \citep{dickey}, which compares the density of
the posterior in the case when the null hypothesis is realised, versus that of
the prior:

\begin{align}
\frac{ \pdf(\mathcal{D}|\Hnull) }{ \pdf(\mathcal{D}|\Hpos) } &= \frac{ \pdf(\mathcal{D}|f=0) }{ \pdf(\mathcal{D}|0<f\leq1) } \nonumber\\
\qquad&= \frac{\lim_{f\to0} \pdf(f|\mathcal{D}) }{\lim_{f\to0} \pdf(f)},
\end{align}

which, using Equation~(\ref{eqn:posterior}), yields

\begin{align}
\underbrace{\frac{ \pdf(\mathcal{D}|\Hpos) }{ \pdf(\mathcal{D}|\Hnull) }}_{\equiv \Kpos} &= 
\frac{ \mathrm{B}_{1-\CPP}[\Ntot-\Npos+1,\Npos+1] - \mathrm{B}_{(1-\CPP)(1-\TPP)}[\Ntot-\Npos+1,\Npos+1] }{ (1-\CPP)^{\Ntot-\Npos+1} \CPP^{\Npos} \TPP },
\label{eqn:Kpos}
\end{align}

where note that we have flipped the ratio of the two hypotheses, such that
it equals the Bayes factor of the positive hypothesis to the null - denoted
more compactly as simply $\Kpos$.  A useful limiting case to note is that

\begin{align}
\lim_{\Ntot \to 0} \Kpos = 1.
\label{eqn:Kzero}
\end{align}

This can be seen by noting that $\mathrm{B}_{x}[1,1] = x$ and thus the
numerator equals $\TPP (1-\CPP)$, which then cancels out with the denominator.
Equation~(\ref{eqn:Kzero}) reveals that with no data, the Bayes factor gives no
preference to either model (as expected).


\section{Classifying the Landscape of Experiments}
\label{sec:classes}

\subsection{A Single Datum Case Study}

A particularly instructive case to consider is to change in our beliefs in
going from no data to one data point. In the absence of any data,
Equation~(\ref{eqn:Kzero}) dictates that the odds ratio between the two
hypotheses simply equals whatever our \textit{a-priori} belief is. Accordingly,
the ratio of the odds factor before/after conducting a single experiment must
equal $\lim_{\Ntot \to 1} \Kpos$ (i.e. the model prior cancels). This single
term thus describes how much we learnt through the act of conducting a single
experiment versus no data. In this way, we can see that $\lim_{\Ntot \to 1}
\Kpos$ is clearly a term we wish to maximise in designing our experiment.

\begin{figure*}
\gridline{\fig{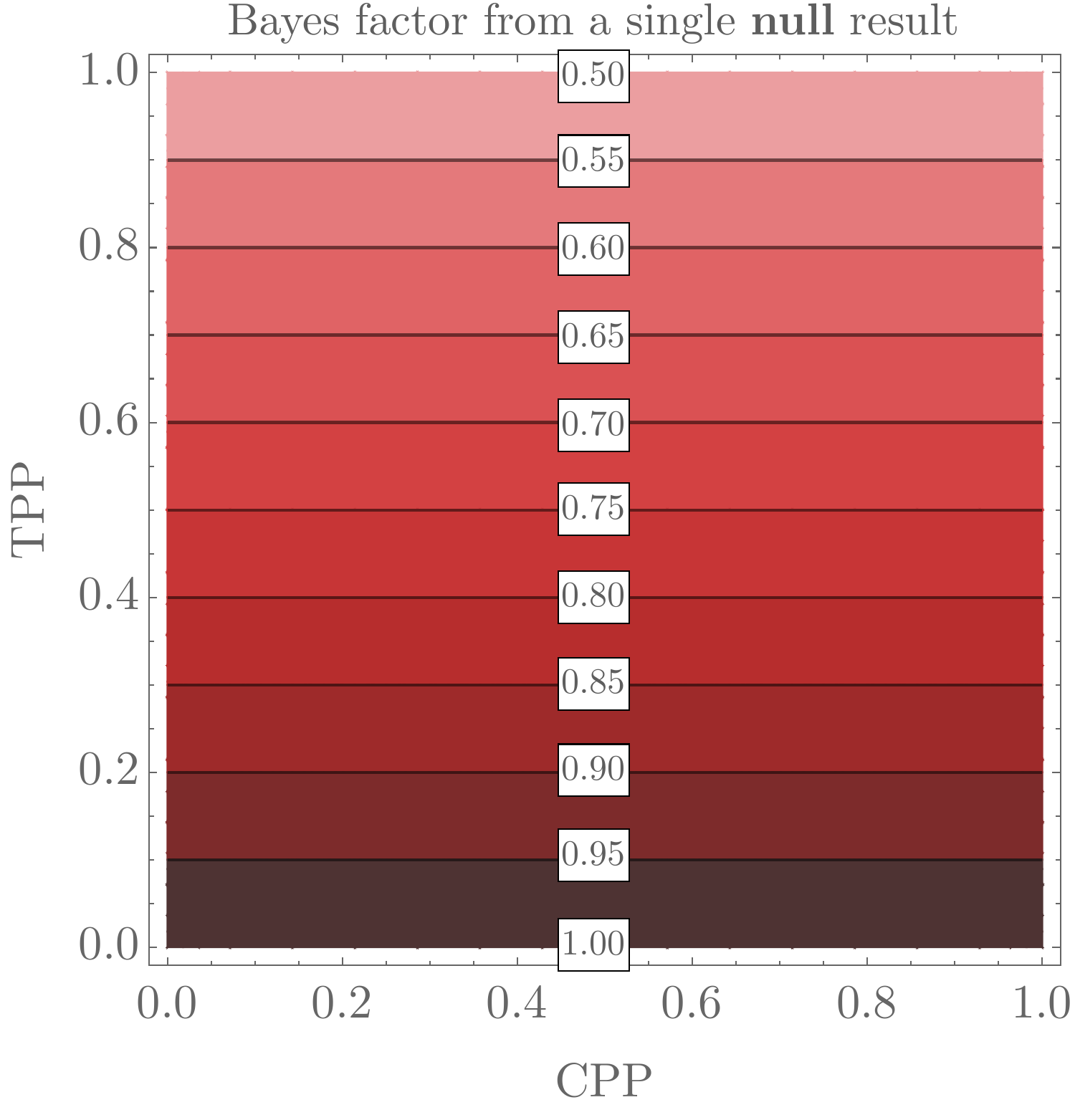}{0.45\textwidth}{(a)}
          \fig{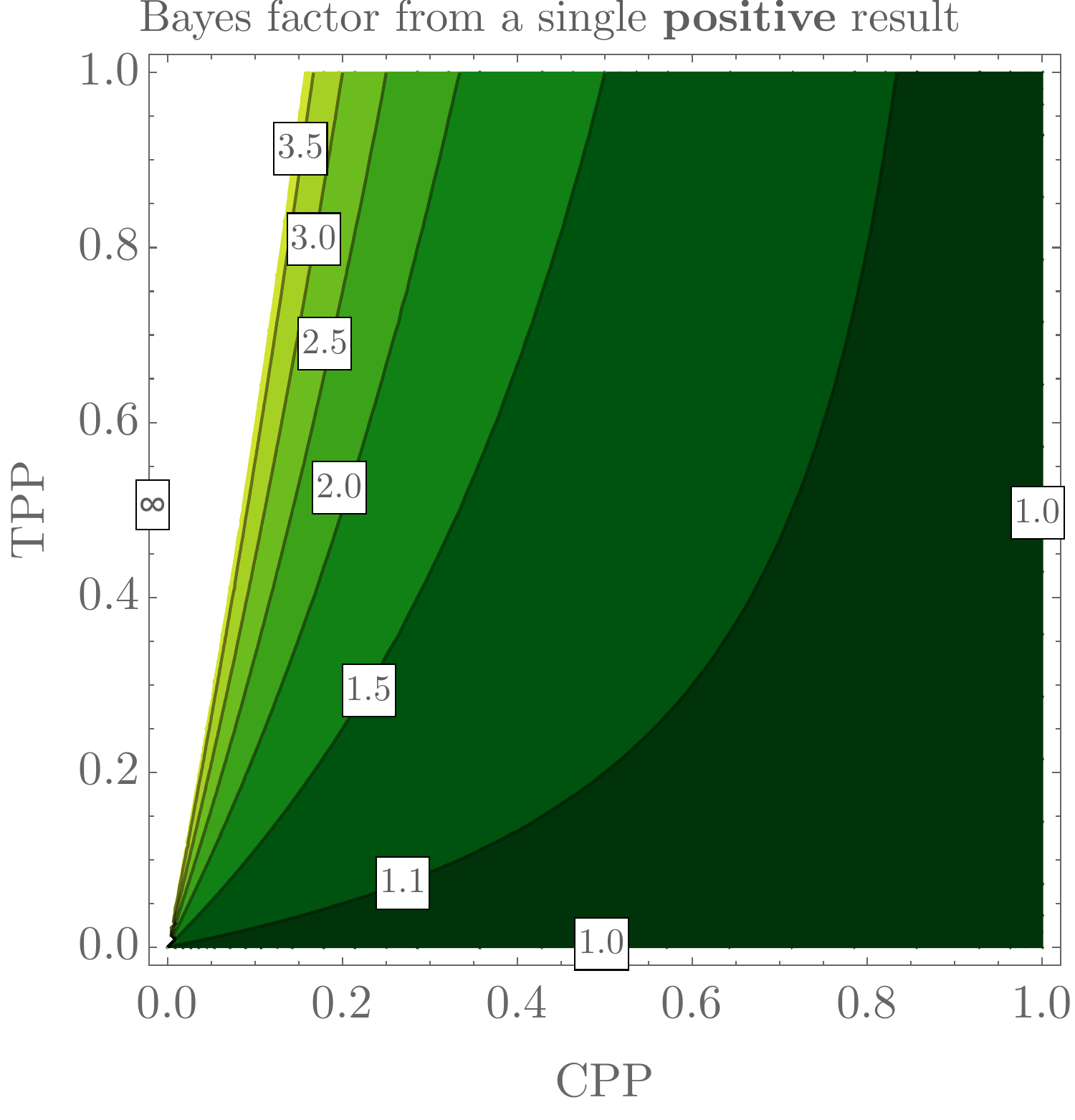}{0.45\textwidth}{(b)}
          }
\caption{
Examples showing the behaviour of the Bayes factor for a single datum,
$\lim_{\Ntot\to1}\Kpos$ (Equation~\ref{eqn:Kone}), assuming our datum is a null
detection (a), or a positive detection (b).
\label{fig:Koneexamples}}
\end{figure*}

In this case of $\Ntot=1$, only two possible observational states exist; either
$\Npos=0$ or $1$. In such a simplifying case, the Bayes factor (also plotted
in Figure~\ref{fig:Koneexamples}) becomes

\begin{equation}
\lim_{N_{\mathrm{tot}} \to 1} \Kpos =
\begin{cases}
1 - \frac{\TPP}{2}  & \text{if } \Npos = 0 ,\\
1 + \frac{\TPP}{2} \Big( \frac{1}{\CPP} - 1 \Big)  & \text{if } \Npos = 1.
\end{cases}
\label{eqn:Kone}
\end{equation}

It is instructive to evaluate Equation~(\ref{eqn:Kone}) in the case of an ideal
experiment, i.e. that $\TPP=1$ and $\CPP=0$. Notably, this leads to the result
that $\Kpos = \infty$ when $\Npos=1$. At first, this might seem
counter-intuitive - how can one possibly have absolute confidence in the
positive hypothesis using just a single datum? The answer is that in an ideal
experiment such as this, it's simply impossible to produce a positive result
unless the positive hypothesis is true. In particular, since $\CPP=0$, there is
zero probability of a given positive being a confounding positive - it
\textit{must} be real. The reason why this may feel counter-intuitive is that
an ideal experiment is of course unattainable in real world conditions.

If, instead, a null result is obtained with the ideal experiment, then one 
finds $\Kpos = \tfrac{1}{2}$. Here, even with our ideal experiment, the result
applies only modest pressure to the $f>0$ hypothesis. How can we understand
this, given the previous case? The answer lies in the fact that with
intermediate values of $f$, sometimes the phenomenon is operating and sometimes
it is not. Thus, we are perhaps here simply observing one of the cases where it
does not. Since we adopted a uniform prior in $f$, for which the expectation
value is $\tfrac{1}{2}$, then this explains why the marginalised likelihood
ratio lands on $\tfrac{1}{2}$ - much like flipping a fair coin.

\subsection{Minimum Experimental Requirements for a Falsifiable/Verifiable Hypothesis}
\label{sub:falsifiable}

It is interesting to note that in the case of $\Ntot\geq1$ and a series of null
results from an ideal experiment, Equation~(\ref{eqn:Kpos}) becomes $\Kpos \to
(\Ntot+1)^{-1}$. This allows us to set a useful limit. Consider that we set a
threshold Bayes factor of $K_{\mathrm{crit}} = 10$ to define ``strong
evidence'' in favour of one of the competing hypotheses\footnote{It is indeed quite typical for surveys/projects/disciplines to adopt statistical thresholds like this; for example the particle physics community often 5\,$\sigma$ \citep{lyons:2013}.}, following the \citet{kass} scale. Such an approach
would be appropriate if there is no \textit{a-priori} preference between the
competing hypotheses - a situation we will here label as the ``agnostic
experimenter'' (i.e. $\pdf(\Hpos)/\pdf(\Hnull)=1$). In this case, a series of
null results from even an ideal experiment would need to amass at least $\Ntot
\geq 9$ targets to reach this threshold, at which point
$\pdf(\Hnull|\mathcal{D}) \geq 10 \pdf(\Hpos|\mathcal{D})$. Thus, in
order for an experimental setup to be \textit{falsifiable}, a cornerstone of the
scientific process \citep{popper}, at least 9 experiments would be required
(more generally one needs $\geq(K_{\mathrm{crit}}-1)$ experiments). Since
an ideal experiment can achieve $\Kpos \to \infty$ from just a single
detection, then we require $\Ntot \geq 1$ for an ideal experiment to be
\textit{verifiable}.

\subsection{Impotent Experiments}

Returning to Equation~(\ref{eqn:Kone}), one can also see that if $\TPP \to 0$,
$\Kpos$ tends to unity. Whilst this is clearest by inspection of
Equation~(\ref{eqn:Kone}), the same statement is true of the more general form
for $\Ntot\geq1$ experiments shown in Equation~(\ref{eqn:Kpos}). Since the
information gain in Shannon entropy bits equals $\log_2 \Kpos$, such
experiments quite literally provide no information whatsoever about the
relative probability of the competing hypotheses, \textit{irrespective} of the
data collected. Such experiments are thus acts of futility, and we label them
in what follows as ``impotent experiments''.

From Equation~(\ref{eqn:Kone}), one can see that a similar situation occurs if
$\Npos=1$ and $\CPP=1$, yielding $\Kpos=1$. One might counter that the
experiment here is not truly impotent because $\Kpos\neq1$ in the case of
$\Npos=0$ (which in fact has no dependency on $\CPP$). However, recall that if
the $\CPP=1$ then every experiment returns a positive and thus the case of
$\Npos=0$ is unrealisable. Thus, such an experiment is similarly doomed to
always return $\Kpos=1$.

Together then, we classify an ``ideal experiment'' as one with $\TPP=1$ and
$\CPP=0$, an ``imperfect experiment'' as one with $0 < \TPP \leq 1$ and
$0 \leq \CPP < 1$, and an ``impotent experiment'' as one with $\TPP=0$
\textit{or} $\CPP=1$.

\subsection{The Influence of the Model Prior}

\noindent\textcolor{quotecol}{Extraordinary claims require extraordinary evidence}\\
Carl Sagan\\

Recall from Equation~(\ref{eqn:oddsratio}) that the ratio of our belief
in the affirmative versus the null hypothesis conditioned upon the data,
$\pdf(\Hpos|\mathcal{D})/\pdf(\Hnull|\mathcal{D})$ equals $\Kpos$
multiplied by the ratio of our \textit{a-priori} belief
$\pdf(\Hpos)/\pdf(\Hnull)$. In this way, $\Kpos$ can be seen to
represent the ratio of our final hypothesis odds-ratio, to our
initial hypothesis odds-ratio; it represents what we \textit{learned}
from the experiment, since rigorously:

\begin{align}
\Kpos &= \frac{\pdf(\mathcal{D}|\Hpos)}{\pdf(\mathcal{D}|\Hnull)} = \frac{ \frac{\pdf(\mathcal{D}|\Hpos)}{\pdf(\mathcal{D}|\Hnull)} \frac{\pdf(\Hpos)}{\pdf(\Hnull)} }{ \frac{\pdf(\Hpos)}{\pdf(\Hnull)} } = \frac{ \overbrace{\frac{\pdf(\Hpos|\mathcal{D})}{\pdf(\Hnull|\mathcal{D})}}^{\mathrm{\textit{a-posteriori}\,\,belief\,\,ratio}} }{ \underbrace{\frac{\pdf(\Hpos)}{\pdf(\Hnull)}}_{\mathrm{\textit{a-priori}\,\,belief\,\,ratio}} }.
\end{align}

Further, the model priors are not necessarily priors guessed arbitrarily, but
could be informed from $N$ previous experiments (i.e.
$\pdf(\Hpos)/\pdf(\Hnull)$ would equal the \textit{a-posteriori} belief ratio
conditioned upon the $N$ previous experiments). Rigorously then, whilst $\Kpos$
was derived for the toy case of a single experiment, it represents (or really
$\log\Kpos$) the information gain in going from $N$ to $N+1$ experiments and
thus must generically be able to deviate from unity in order for us to have a
useful experiment of any kind.

Whilst focussing on information gain before/after an experiment dissolves
the model prior, the prior ratio can play an important role in considering
other interpretative measures which can be illustrated with an extreme,
limiting example. For ideal or imperfect experiments with $\TPP >0$ and
$\CPP<1$, we've seen how $0<\Kpos\leq\infty$. This means that whilst we could
achieve $\pdf(\mathcal{D}|\Hpos)/\pdf(\mathcal{D}|\Hnull) = \infty$
(i.e. $\Kpos=\infty$), we can never achieve
$\pdf(\mathcal{D}|\Hnull)/\pdf(\mathcal{D}|\Hpos) = \infty$
(i.e. $\Knull=(1/\Kpos)=\infty$). In other words, we could end up with
absolute belief in the affirmative hypothesis, but can never reach
such a state for the null. Put simply, one can prove something exists,
but one can never prove (absolutely) it does not exist.

This subtle point has an important consequence. If the experimenter approaches
their task with $\pdf(\Hpos)/\pdf(\Hnull) \to \infty$ as their prior belief,
then since $\Knull < \infty$, even an enormous amount of null data would be
unable to overturn the experimenter's prior. We here label such as
experimenter as a ``dogmatic experimenter'' - which juxtaposes the agnostic
experimenter introduced in Section~\ref{sub:falsifiable}. Likewise,
a dogmatic experimenter could hold the opposite extreme of
$\pdf(\Hpos)/\pdf(\Hnull) = 0$, in which case no imperfect experiment(s)
can ever convince them that the affirmative hypothesis is correct.
Rigorously, there's no point in the dogmatic experimenter doing any
experiments since they cannot learn anything from the exercise (and indeed
we advocate they do not waste their time in doing so).

Leaving aside the clearly extreme and limiting case of
$\pdf(\Hpos)/\pdf(\Hnull) = \infty$, consider now the softer case of
$\pdf(\Hpos)/\pdf(\Hnull) \gg 1$. The appropriateness of
such a strong prior model preference is problem specific. For example, if a
single observer reported that a glass fell up into the sky rather than down,
defying countless experiments/experiences, it would be appropriate to not
immediately reject our theory of gravity. Here, a strong model prior is
justified from a large amount of prior data, which speaks to Sagan's
adage. However, we would generally expect that in a resource-constrained
environment experiments are performed to test theories for which considerable
uncertainty exists regarding their validity, rather than those with
overwhelming confidence already. Thus, in practice, it would usually be
inappropriate for an experimenter working on the frontier of human knowledge to
adopt such a strong prior. Nevertheless, personal prejudices can inject heavy
bias into such analyses and demand constant vigilance against
\citep{goldstein:2021}.

In such cases where $\pdf(\Hpos)/\pdf(\Hnull) \gg 1$, the experimenter has
a strong expectation that the positive hypothesis is correct before seeing any
data, and thus we describe them as a ``biased experimenter'' (which despite
the negative connotations of that word, is not necessarily problematic). Recall
that hypotheses are often accepted if they cross some threshold odds ratio; for
example $K_{\mathrm{crit}}=10$ represents ``strong evidence'' on the
\citet{kass} scale. Naturally then, if our prior ratio is
overwhelming strong (i.e. $\pdf(\Hpos)/\pdf(\Hnull) \gg K_{\mathrm{crit}}$),
then we require very strong evidence for the experiment to overturn the prior
belief.

Adopting a non-unity prior ratio affects the calculation presented in
Section~\ref{sub:falsifiable}, which previously explicitly assumed an agnostic
experimenter. In particular, consider the case of a series of null results
which should apply pressure on the positive hypothesis for imperfect/ideal
experimental designs. Including a generalised prior ratio, we now find that
an ideal/imperfect (=/$>$) experiment would require

\begin{align}
\Ntot \geq \frac{\pdf(\Hpos)}{\pdf(\Hnull)} K_{\mathrm{crit}} - 1
\end{align}

targets all returning null results to reject the positive hypothesis i.e. to be
falsifiable at the required statistical threshold. This reveals how our prior
hypothesis assumptions have a direct impact on the experimental design, if one
wishes to retain a falsifiable experiment.

\section{Experimental Design for Alien Hunters}

\subsection{An Initial Experiment}

We have thus far presented a rigorous, analytic framework for interpreting
data emerging from a generalised experiment. We now seek to bridge
this framework to the specific case of looking for alien life. For the
moment, we endeavour to maintain a broad scope here, with experiments spanning
everything from looking for life in the shadow biosphere (e.g.
\citealt{benner}) to alien radio transmissions (e.g. \citealt{cocconi}), from
exoplanet biosignatures (e.g. \citealt{seager:2005}) to seeking alien artefacts
within the Solar System (e.g. \citealt{bracewell}).

We require the hypotheses to be diametrically opposed such that their
prior probabilities sum to unity. We begin with the simplest version of the
hypotheses one might consider, but will re-evaluate them as we proceed:

\begin{itemize}
\item[{$\Haliens$}] Alien life exists within some non-zero fraction $f$ of the
sample observations.
\item[{$\overline{\Haliens}$}] Alien life does not exist within any of the observed sample
(i.e. $f=0$).
\end{itemize}

In the above, ``aliens'' is defined as life, or entities derivative from life,
that stems from a distinct abiogenesis event to that of ourselves, but could be
any level of evolutionary complexity.

To make progress, we will propose an initial experiment (which we'll dub
``Experiment A''), consider its usefulness, and then iteratively refine as
needed. Although the experimental labelling scheme will be explicitly defined
in what follows, the data in question is left deliberately general to maximise
the scope of this study. Data specific case examples are discussed later in
Section~\ref{sec:discussion}.

The experimental labelling schemes corresponding to our series of iterative
designs are presented in full in Table~\ref{tab:experiments}. The experimenter
classifies experimental data, $\mathcal{D}$, as either a detection
($\mathcal{S}$) or as a null-detection ($\bar{\mathcal{S}}$), following the
system provided in that table. 

\begin{table}[h!]
\caption{
Three possible schemes that could be used to define a success ($\mathcal{S}$) 
and null ($\bar{\mathcal{S}}$) label. Each defines a unique experiment,
and they are treated in the text as following iteratively from the previous.
The term ``combigenic'' is defined in Section~\ref{sub:INU}, and includes
abiogenic (not originating from living systems) and lucagenic (originating
from life derioved from LUCA).
} 
\centering 
\begin{tabular}{c c c} 
\hline\hline
Experiment & $\mathcal{D}=\mathcal{S}$ & $\mathcal{D}=\bar{\mathcal{S}}$ \\ [0.5ex] 
\hline
A & Data can be explained with & Data cannot be explained by  \\
& alien biology/behaviour/technology &alien biology/behaviour/technology \\
\hline
B & Data is inconsistent & Data is consistent \\
& with combigenic processes & with combigenic processes \\
\hline
C & Data is inconsistent & Data is consistent \\
& with known combigenic processes & with known combigenic processes \\
\hline 
\end{tabular}
\label{tab:experiments} 
\end{table}

For the moment, we direct the reader to just the first item of that table -
Experiment A - which we will initially adopt as a starting point.

\newpage
\subsection{Alien Headaches}

Before discussing the labelling schemes in depth, we first pause to crystallise
exactly what it is about alien hunting that poses such a challenge to
conventional scientific experimentation. It's easy to see that ``aliens'', as a
broadly defined concept, raise several problems atypical of conventional
scientific experiments. Notably, we do not know whether they exist at all,
we lack any prior data on them, and they may operate with unpredictable
agency/limits. Stemming from this, we identity three crucial features that
would potentially compromise experimental efforts to detect them:

\begin{enumerate}
\item \underline{Unbounded Explanatory Capacity (UEC):}
A positive detection of \textit{any} phenomenon can be explained with
sufficiently advanced aliens, since we do not know their capabilities or
behaviour.
\item \underline{Unbounded Avoidance Capacity (UAC):}
A null result from experiments seeking aliens can \textit{always} be explained
as consistent with the behaviour of sufficiently elusive and/or advanced aliens.
\item \underline{Incomplete Natural Understanding (INU):}
Positive detections of aliens reasoned through the deductive exclusion of all
explanations devoid of biology/technology may be spurious, since our
understanding of such processes is incomplete.
\end{enumerate}

\subsection{Unbounded Explanatory Capacity (UEC)}
\label{sub:UEC}

\noindent\textit{\textcolor{quotecol}{
Any sufficiently advanced technology is indistinguishable from magic}}\\
Arthur C. Clarke\\


UEC is a direct product of our rather vaguely defined (deliberately so)
labelling system. ``Aliens'' as a broad concept can explain just about
\textit{anything} when one admits they may be far more biologically or 
technologically sophisticated than us with no obvious ceiling, thereby
enabling the hypothesis to explain observations ranging from deep field
observations of our cosmos to why your door swings at night. For example,
one could consider that our Universe is the product of alien activity (e.g.
\citealt{bostrom:2003}), or indeed is itself conscious in some form (e.g.
\citealt{hossenfelder:2022}).

The statistical consequence of this is that when one confronts the alien
hypothesis with a positive detection of some phenomenon, indeed \textit{any}
phenomenon, the likelihood function (which recall is the probability of
obtaining the data given the hypothesis) is high and approaches unity.


Formally integrating this feature into our analytic formalism is sensitive
to the labelling scheme used. In Experiment A, the phenomenon in question is
labeled as a success if it is deemed ``consistent with aliens''. The
dichotomous experimental labelling scheme demands that the null label is thus
``inconsistent with aliens''. The experimenter would now sift through a series
of targets and assign labels appropriately. Of course, if one uses this
labelling criterion and accepts the UEC of aliens, then \textit{all} targets
would be labeled as successes, $\mathcal{S}$, meaning that $\Npos = \Ntot$. A
null label is simply unobtainable. It should be immediately obvious that such
an experiment will be impotent, as the result is always the same irrespective
of the data presented (i.e. the data do not and cannot affect the outcome).

In the instructive case of a single datum, Equation~(\ref{eqn:Kone}) can be
used to see this analytically. Since UEC leads to $\Npos=\Ntot$, then in the
single datum case we have $\Npos=1$ and thus $\Kpos = 1 +
\tfrac{\TPP}{2} (\CPP^{-1}-1)$.

Recall that $\CPP$ is defined as the fraction of all experiments which are
labeled as a positive detection irrespective of which hypothesis is acting.
So here, this would be the fraction of observed phenomena which are classified
as consistent with alien activity ($\mathcal{S}$) irrespective as to whether
aliens exist or not ($\Haliens$ and $\overline{\Haliens}$ respectively). Since
all experiments are labeled as positive via UEC, then we must have $\CPP=1$,
which means that $\Kpos = 1$ and we have an impotent experiment.


\subsection{Unbounded Avoidance Capacity (UAC)}
\label{sub:UAC}

\noindent\textit{\textcolor{quotecol}{
An absence of evidence is not evidence of absence.}}\\
Carl Sagan\\

Another approach to Experiment A is not to focus on positive labels per se,
but rather look for labels which cannot be labeled as negative - following
a \textit{reductio ad absurdum} perspective. This shift in perspective
is instructive to highlight the second challenge plaguing alien hunters - UAC.

A pedagogical example might be to survey Mars for living organisms and report
no obvious signatures of life. Despite this, we cannot label the experiment as
$\bar{\mathcal{S}}$, which recall states that the ``data cannot be explained by
aliens''. In fact, alien life could very well be on Mars, just evading the
sensitivity of the experiment itself. For example, the experiment could be
sampling the surface but life in fact lives deep beneath the surface.

This formally represents a high false negative probability ($\FNP$). Since
$\TPP = 1 - \FNP$, then this is equivalent to a low $\TPP$ case i.e. an
experiment of zero sensitivity. Interpreting a low $\TPP$ within our framework
is straight-forward but depends upon the precise value. If we have truly
unbounded evasion, then $\TPP \to 0$. Irrespective of the $\CPP$, $\Kpos \to
1$ in this limit (e.g. see Equation~\ref{eqn:Kone}) and thus we again have an
impotent experiment. We can thus see that Experiment A is impotent both from
$\CPP\to1$ \textit{and} $\TPP\to0$.

One might argue that our limiting case evaluations of the UEC producing
$\CPP \to 1$ and the UAC producing $\TPP \to 0$ are too extreme. For the
UEC, consider then a softer version where that aliens have \textit{potential}
UEC (i.e. we admit the possibility but don't necessitate it), which in turn
means that all data can \textit{potentially} be explained by aliens. In practice,
any attempt at statistical progress here requires some quantification of how
often/probable/extensive of this potentiality, which of course cannot be
reliably assessed in the absence of any alien sample beforehand. Likewise for
UAC, we might consider our experiment to have a low but non-zero $\TPP$.
However, we can't legitimately make statements like $\TPP = 0.01$, since it
depends upon their evasiveness - which is of course \textit{a-priori} unknown!
Any $\TPP$ we adopt would be conditional upon some specific assumptions about
the abilities, agency and appearance of alien life, which is in general unknown
prior to their detection. Both the UEC and UAC problems underscore why we
consider Experiment A a highly problematic way of attempting to search for
alien life and some revision is clearly needed.

\subsection{Incomplete Natural Understanding (INU)}
\label{sub:INU}


\noindent\textit{\textcolor{quotecol}{
There are known knowns; there are things we know we know...
But there are also unknown unknowns, the ones we don't know we don't know}}\\
Donald Rumsfeld \\

Both UEC and UAC concern the behaviour/activity of alien life. Since that
behaviour is \textit{a-priori} unknown, Experiment A is ineffectual as a
result. In this way, we can see that our next iteration should seek to frame
the labelling system not in terms in \textit{what aliens do}, but rather
\textit{what a universe devoid of aliens cannot do}.

The experimental approach here adopts a kind of Sherlock Holmes philosophy,
attempting to establish the alien hypothesis by elimination of the
alternatives. In such an experiment, one can see that all it takes for the
alien hypothesis to triumph is for the null hypothesis to fail. This already
sets up a precarious scientific enterprise though; the broadest version
of the alien hypothesis can never fail to explain the data to high likelihood
(via UEC), but the alternatives occasionally will, simply as a result of
probabilistic sampling. For in any sufficiently large sample of observations,
there will inevitably be outlier measurements that are deemed improbable via
the natural hypothesis, and it is in those moments that the alien hypothesis
can ostensibly enjoy elevation to a credible solution. This situation
resembles the publication bias component of the replication crisis plaguing
many sciences today \citep{francis:2012}, which can also be driven by
outlier events.

Ideally, this would be resolved through the accurate modelling of outliers
under the null hypothesis. Afterall, if our natural understanding is total,
then one should be able to correctly predict the number of outliers. Such an
experimental design puts enormous pressure on our knowledge of the mechanics
of our universe though, since we're demanding it provides a highly accurate
prediction of what essentially constitutes the ``tails'' of probability
space. We must admit the possibility that our knowledge may be simply
incomplete in these tails.

Within astronomy, there is certainly plenty of historical examples of the alien
hypothesis being put forward when confronted with observations not anticipated
by the prevailing astrophysical theories. These range from early observations of
Mars leading to claims of a canal system \citep{lowell:1906}, to more recent
claims concerning alien origins for fast radio bursts \citep{lingam:2017} and
interstellar asteroids \citep{loeb:2022}. In many ways, the alien hypothesis
here parallels the ``God of the Gaps'' argument familiar to theologians
\citep{heller:1993}.

The above examples speak to astronomy - hence in Experiment B one would
seek observations forbidden by astrophysics. Already one might see a problem
though; that one could reasonably argue that alien life is a legitimate part of
astrophysics, and indeed this linguistic problem is exacerbated when one steps
back out of astronomy to science more broadly. Words like natural and unnatural
are also inaccurate, since even technological mega-structures like Dyson
spheres \citep{dyson:1960} could be argued to be ultimately natural.

A more nuanced classification that Experiment B endeavours to achieve involves
categorising a given phenomenon into one of two fundamental categories. The
first category encapsulates phenomena explainable by \textit{abiogenic}
physical/chemical processes, where neither living systems nor their derivatives
(such as technology) play a role. Alternatively, this category also includes
phenomena resultant from life that shares a common origin with humanity in the
abiogenesis event, termed as \textit{``lucagenic''}. Here, the
prefix ``luca'' subtly nods to the Last Universal Common Ancestor, LUCA
\citep{woese:1990}. The second category is demarcated as the mutual exclusive
of the first. We choose to refer to this as \textit{``allogenic''}, where
``allo'' serves as a prefix symbolising ``otherness''.

Navigating through terminological choices, a synthesis is essential for concise
notation and communication. Merging ``abiogenic'' and ``lucagenic'' presents a
terminological challenge due to the inherent diversity and complexity of the
concepts involved. No solitary prefix impeccably encapsulates the collective
essence of these terms when standing alone. Thus, we gravitate towards an
abstract yet illustrative term, \textit{``combigenic''}, to symbolise the 
integrative nature of these two foundational categories.

With this discussion in mind, let us now try to integrate the modified
experimental setup into our formal framework. Let us turn to Experiment B,
with the modified labeled scheme presented in Table~\ref{tab:experiments}.
Now, a positive label is assigned to data classified as allogenic. A subtlety
here is that unlike Experiment A, $\mathcal{S}$ no longer has direct
connotation to $\Haliens$. Specifically, in Experiment B, such detections do
not strictly translate to supporting the alien hypothesis, since in principle
one might propose that some alternative explanation exists.

Technically, then, we should write that a positive label here directly
supports hypothesis $\unnatur$ (rather than $\Haliens$). However, if we
continue under the simplifying (and wholly reasonable) naturalistic assumption 
that supernatural entities, deities, etc do not exist, then by deduction
allogenic events can only be caused by aliens and thus now directly connect
back to our original hypotheses, such that $\pdf(\Haliens|\unnatur) = 1$.

With this point aside, an outstanding challenge with Experiment B is that our
knowledge of lucagenic, and especially abiogenic, processes is incomplete.
Thus, we risk labelling some events as detections when they are in fact merely
a consequence of some undiscovered physical process(es), for example. Such
spurious detections contribute to the list of confounding positives, but the
rate at which they occur cannot be assigned a firm value, since they concern as
yet unknown processes. Accordingly, the $\CPP$ can also not be easily
quantified and becomes fuzzy.

With fuzzy parameters like this, a common remedy is to marginalise over $\CPP$,
but this requires an assumed prior distribution for $\Pr(\CPP)$. A simple
approach might be be to adopt a uniform distribution from $\CPP_{\mathrm{min}}$
to 1, where $\CPP_{\mathrm{min}}$ essentially corresponds to the case where
new processes never operates and reflects the experiment's inherent confounding
rate. However, this gives equal weight to all possible $\CPP$ values for which
we have no justification to suspect is reasonable. For all we know, previously
unknown processes might be a very frequent source of confounding positives and
thus the marginalised $\Kpos$ would be a wild overestimate - indeed the history
of spurious alien claims in science adds credence to this possibility (e.g.
\citealt{lowell:1906}, \citealt{hewish:1968}, \citealt{wright:2016}). This is
obviously going to depend on the precise context and type of observation and
phenomenon being studied. However, in general, the unknown nature of how
often/likely new processes crops up within the data clearly obstructs the
analysis, but this probability can be put front and centre more explicitly
through a final iteration of our experimental design to Experiment C.

\subsection{Separating Out the Alien Question}
\label{sub:exptC}

Experiment C limits positive detections to those inconsistent with
\textit{known} combigenic processes. But this re-framing comes at a cost.
Recall that for Experiment B, the alternative hypothesis is not strictly
``aliens'', but also technically includes unnatural phenomena. However, in the
absence of the supernatural, there's no credible alternative but to invoke
aliens in such a case and thus we can (and did) equate them. However, in
Experiment C, the alternative hypothesis is strictly a phenomenon
inconsistent with \textit{known} combigenic (=lucagenic or abiogenic)
processes ($\unnaturX$). This change is crucial, since it expands the causal
explanations from primarily aliens to now aliens plus as-yet-unknown processes
(which is generally quite plausible).

\begin{figure*}
\begin{center}
\includegraphics[width=15.0cm,angle=0,clip=true]{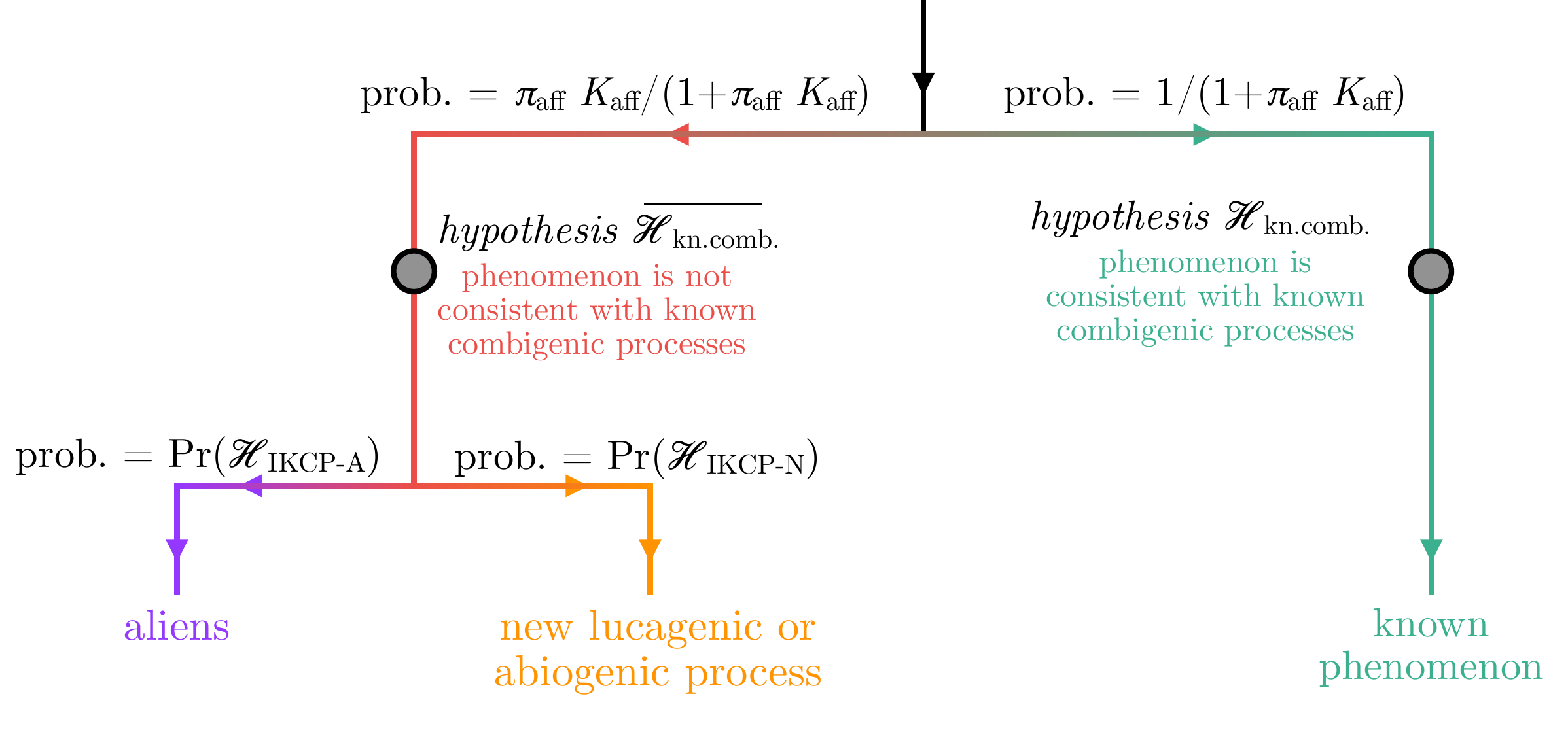}
\caption{
Schematic of the branching points in our model comparison framework for
Experiment C. The primary hypothesis is whether the observed phenomenon is
consistent (right branch) or inconsistent (left branch) with \textbf{known}
combigenic phenomena. However, the left branch does not yet establish the alien
hypothesis, since two competing sub-hypotheses exist, as shown.
}
\label{fig:subtree}
\end{center}
\end{figure*}

In Experiment B, if the affirmative hypothesis were to be established
($\unnatur$), then there's little plausible alternative but to equate this to
accepting our original alien hypothesis (i.e. $\pdf(\Haliens|\unnatur)=1$).
But in Experiment C, establishing the affirmative hypothesis is not enough - as
now competitive sub-hypotheses exist (i.e.
$\pdf(\Haliens|\unnaturX)\leq1$). Strictly, $\Kpos$ now represents
$\pdf(\mathcal{D}|\unnaturX)/\pdf(\mathcal{D}|\naturX)$. This branching into
sub-hypotheses is depicted in Figure~\ref{fig:subtree}.

On the one hand, one might be satisfied with this as stated. If we follow
Experiment C and obtain $\Kpos \gg 1$, the we have discovered something
profound either way - a previously unknown processes or alien life! Indeed,
this has long been the appeal of SETI\footnote{See \href{https://www.seti.org/could-we-really-find-et}{https://www.seti.org/could-we-really-find-et}.}.
The experimenter can perhaps be content with this, and indeed history
is littered with such detections that were initially somewhat ambiguous
as to whether they were caused by aliens or new physics, such as PSR B1919+21
\citep{hewish:1968} or Boyajian's Star \citep{boyajian:2016,wright:2016}.

\subsection{Model Comparison in Experiment C}

For those seeking the alien question directly, the ambiguity establishes
that model comparison will be necessary to directly interrogate the alien
hypothesis against alternatives. Let us group these alternatives into a
single broad hypothesis: a phenomenon that is inconsistent with known
combigenic processes and yet is nonetheless either lucagenic or abiogenic
i.e. a new (previously undiscovered) combigenic phenomenon ($\IKNPn$ =
Inconsistent with Known Combigenic Processes and yet Combigenic). Given
the Boolean construction of that statement, we have
$\pdf(\IKNPn) = \pdf(\natur,\unnaturX)$.

Diametrically opposed to this, we have the hypothesis of $\IKNPa$: a phenomenon
that is both inconsistent with known combigenic processes and is caused by
aliens, such that $\pdf(\IKNPa) = \pdf(\Haliens,\unnaturX)$. Since these
two alternatives are mutually exclusive and entail all possibilities, then
$\IKNPn = \olsi{\IKNPa}$.

Whilst we appreciate that the terminology becomes rather cumbersome, it's
important for specificity as one can more easily identify terms that should not
be conflated. In particular, $\pdf(\IKNPa) \neq \pdf(\Haliens)$ since
$\pdf(\IKNPa) = \pdf(\unnaturX|\Haliens) \pdf(\Haliens)$ and one cannot
assume that the activities of aliens would necessarily violate known
combigenic processes ($\pdf(\unnaturX|\Haliens) \neq 1$).

Consider that we wish to solve for the odds ratio, $\mathcal{O}$, given by

\begin{align}
\mathcal{O} &= \frac{ \pdf(\IKNPa|\mathcal{D}) }{ \pdf(\olsi{\IKNPa}|\mathcal{D}) }.\\
\qquad&= \frac{\pdf(\text{\scriptsize{phenomenon\,\,is\,\,inconsistent\,\,with\,\,known\,\,abiogenic/lunagenic\,\,processes\,\,and\,\,is\,\,caused\,\,by\,\,aliens}}|\text{\scriptsize{data}})}{\pdf(\text{\scriptsize{phenomenon\,\,is\,\,inconsistent\,\,with\,\,known\,\,abiogenic/lunagenic\,\,processes\,\,and\,\,is\,\,not\,\,caused\,\,by\,\,aliens}}|\text{\scriptsize{data}})}\nonumber
\end{align}

One can expand the denominator by the law of sum of probabilities, to give

\begin{align}
\pdf(\olsi{\IKNPa}|\mathcal{D}) =& \pdf(\olsi{\IKNPa}|\unnaturX,\mathcal{D}) \pdf(\unnaturX|\mathcal{D}) \nonumber\\
\qquad& + \pdf(\olsi{\IKNPa}|\naturX,\mathcal{D}) \pdf(\naturX|\mathcal{D}).
\end{align}

We next exploit that, by definition, $\pdf(\olsi{\IKNPa}) = 1 - \pdf(\IKNPa)$, to give

\begin{align}
\pdf(\olsi{\IKNPa}|\mathcal{D}) =& \big[ 1 - \overbrace{\pdf(\IKNPa|\unnaturX,\mathcal{D})}^{=\pdf(\Haliens,\unnaturX|\unnaturX,\mathcal{D})} \big] \pdf(\unnaturX|\mathcal{D}) \nonumber\\
\qquad& + \big[1-\underbrace{\pdf(\IKNPa|\naturX,\mathcal{D})}_{=0}\big] \pdf(\naturX|\mathcal{D}),
\end{align}

where we annotate on two replacements; the first exploiting the definition
$\pdf(\IKNPa)=\pdf(\Haliens,\unnaturX)$ and the second exploiting the fact
that we cannot have an process inconsistent with known combigenic processes
whilst also paradoxically being consistent with known combigenic processes.
Further, since $\pdf(A,B|B,D)= \pdf(A|B,D) \pdf(B|B,D) = \pdf(A|B,D)$, then
that first annotated term further simplifies to give

\begin{align}
\pdf(\olsi{\IKNPa}|\mathcal{D}) =& \big[1-\pdf(\Haliens|\unnaturX,\mathcal{D}) \big] \pdf(\unnaturX|\mathcal{D}) + \pdf(\naturX|\mathcal{D}).
\end{align}

Consider that $\natur$ and $\unnatur(=\mathcal{H}_{\mathrm{allogenic}})$ are
mutually exclusive; all phenomena are truly either combigenic or allogenic.
Thus, $\pdf(\natur) = 1 - \pdf(\mathcal{H}_{\mathrm{allogenic}})$. Further, if
one is willing to accept that the only possible example of an allogenic
phenomenon is alien life or its products (i.e. we discount the supernatural),
then $\pdf(\mathcal{H}_{\mathrm{allogenic}}) = \pdf(\Haliens)$ and hence
$\pdf(\natur) = 1 - \pdf(\Haliens)$. Accordingly, we can use to further
simplify the denominator to


\begin{align}
\pdf(\olsi{\IKNPa}|\mathcal{D}) =& \pdf(\natur|\unnaturX,\mathcal{D}) \pdf(\unnaturX|\mathcal{D}) + \pdf(\naturX|\mathcal{D}),
\end{align}

which now an odds ratio of

\begin{align}
\mathcal{O} &= \frac{ \pdf(\Haliens,\unnaturX|\mathcal{D}) }{ \pdf(\natur|\unnaturX,\mathcal{D}) \pdf(\unnaturX|\mathcal{D}) + \pdf(\naturX|\mathcal{D}) }\nonumber\\
\qquad&= \frac{ \pdf(\Haliens|\unnaturX,\mathcal{D}) \pdf(\unnaturX|\mathcal{D}) }{ \pdf(\natur|\unnaturX,\mathcal{D}) \pdf(\unnaturX|\mathcal{D}) + \pdf(\naturX|\mathcal{D}) }.
\end{align}

If we expand out the terms $\pdf(\naturX|\mathcal{D})$ and
$\pdf(\unnaturX|\mathcal{D})$ with Bayes' theorem, with some
re-arrangement we find

\begin{align}
\mathcal{O} &= \frac{ \pi_{\mathrm{aff}} \Kpos \pdf(\Haliens|\unnaturX,\mathcal{D}) }{
\pi_{\mathrm{aff}} \Kpos \pdf(\natur|\unnaturX,\mathcal{D}) + 1,
}
\end{align}

where we have made the substitutions
$\Kpos = \pdf(\mathcal{D}|\unnaturX)/\pdf(\mathcal{D}|\naturX)$
and $\pi_{\mathrm{aff}} = \pdf(\unnaturX)/\pdf(\naturX)$. Finally, we
note that the data $\mathcal{D}$ only speaks to the hypotheses
$\naturX$ and $\unnaturX$ and is assumed have no discriminatory
power (i.e. information pertaining to) the probability of
aliens versus some previously unknown combigenic phenomenon. Accordingly,
the $\mathcal{D}$ conditional remaining in this expression dissolves to
give


\begin{align}
\mathcal{O} &= \frac{ 1-\pdf(\natur|\unnaturX) }{
 \underbrace{\pdf(\natur|\unnaturX)}_{\text{model\,\,prior}} + \big[ \underbrace{\pi_{\mathrm{aff}}}_{\substack{\text{signal} \\ \text{prior}}} \underbrace{\Kpos}_{\substack{\text{signal} \\ \text{Bayes} \\ \text{factor}}} \big]^{-1},
}
\end{align}

and the term $\pdf(\natur|\unnaturX)$ can be understood to be our prior
belief concerning the probability that a previously undiscovered combigenic
phenomenon could explain that which appears inconsistent with known combigenic
processes.

\newpage
\subsection{Limiting Cases}

To gain some intuition from this formula, consider the limiting case
where $\pdf(\natur|\unnaturX)\to0$, which means that an unknown combigenic
process is impossible (i.e. we have complete domain understanding), thus
matching Experiment B. In this limit, we have

\begin{align}
\lim_{\pdf(\natur|\unnaturX)\to0} \mathcal{O} &= \pi_{\mathrm{aff}} \Kpos,
\end{align}

as expected, since this matches Experiment B where there's no concept of
unknown processes. In the opposing limit of $\pdf(\natur|\unnaturX)\to1$,
one obtains

\begin{align}
\lim_{\pdf(\natur|\unnaturX)\to0} \mathcal{O} &= 0,
\end{align}

which can be understood by the fact the prior now demands ``it's never
aliens'' (and thus always some previously undiscovered combigenic phenomenon).
It's interesting to note that in the limit of 
$\pdf(\natur|\unnaturX)\to\pdf(\Haliens|\unnaturX)\to\tfrac{1}{2}$
(i.e. even prior odds) and further that $\Kpos\to1$ and
$\pi_{\mathrm{aff}}\to1$, we obtain

\begin{align}
\lim_{\pdf(\natur|\unnaturX)\to1/2} \lim_{\Kpos\to1} \lim_{\pi_{\mathrm{aff}}\to1} \mathcal{O} &= \frac{1}{3}.
\end{align}

This can be understood by the fact that in this totally agnostic scenario,
there are three equally weighted possibilities; a known combigenic process,
an unknown combigenic process or aliens.

Perhaps the most instructive case to consider is the general case, but where
one demands that $\mathcal{O}>\mathcal{O}_{\mathrm{crit}}$ as a significance
threshold to claim discovery (e.g. $\mathcal{O}_{\mathrm{crit}}=10$ on the
\citet{kass} scale constitutes ``strong evidence''). Re-arranging this
inequality, we find that

\begin{align}
\pdf(\natur|\unnaturX) < \frac{ \pi_{\mathrm{aff}} \Kpos - \mathcal{O}_{\mathrm{crit}} }{
\pi_{\mathrm{aff}} \Kpos (1+\mathcal{O}_{\mathrm{crit}}) }.
\end{align}

Now take the limiting case of this expression where either $\pi_{\mathrm{aff}} \to \infty$
or $\Kpos \to \infty$, corresponding to the case where the experimenter deems absolute
confidence that they have observed an event inconsistent with known combigenic phenomena
(which is clearly a best case scenario). In this limit, in order to claim a significant
discovery, we can deduce a firm limit that must be true and we label as:

\newpage
\begin{center}
\textbf{The Alien Hunter's Criterion}
\end{center}
\vspace{-0.5cm}
\begin{align}
\pdf(\natur|\unnaturX) < \frac{ 1 }{ 1+\mathcal{O}_{\mathrm{crit}} }.
\label{eqn:criterion}
\end{align}

Finally, this expression now clearly elucidates the situaton.
If the above inequality is not satisfied, \textit{there is no hope of ever
confidently establishing the alien hypothesis}. And this is a formidable
challenge - for it asks us to characterise the \textit{a-priori} probability
that a previously unknown lucagenic or abiogenic phenomenon could be causing
an apparent violation of our present understanding.

As an example, with the \citet{kass} ``strong evidence'' threshold of
$\mathcal{O}_{\mathrm{crit}}=10$, the inequality in
Equation~(\ref{eqn:criterion}) requires $\pdf(\natur|\unnaturX)<9.09$\%.
This represents a rather high level of ``completion'', in terms
of our knowledge of the workings of the Universe, and the suitability
of such a claim will clearly depend upon the context, but to be clear -
any alien claim must implicitly be making such a claim, whether the
claimant realises it or not.

\section{Discussion}
\label{sec:discussion}

\subsection{Summary}

This work has described a generalised Bayesian framework for interpreting the
results of an experiment with two outcomes, a positive or null detection,
in the context of seeking alien life. The classification of a positive
detection is not absolute and will be recovered at some true positive
probability (TPP), reflecting the experimental sensitivity and interpretative
confidence. In addition, spurious detections may occur at some confounding
positive rate (CPP). We show how the Bayes factor of the affirmative hypothesis
over the null can be rigorously derived in terms of these factors, without loss
of generalisation, and is given by Equation~(\ref{eqn:Kone}). Further, we show
how this corresponds to the information gain from before-to-after any given
experiment, independent of prior experiments/assumptions.

Our framework introduces the idea of the impotent experiment, where the Bayes
factor is stubbornly fixed at unity (which occurs for $\TPP=0$ or $\CPP=1$),
the ideal experiment (which occurs for $\TPP=1$ and $\CPP=0$), and the
imperfect experiment (all intermediate cases). We also describe how
\textit{a-priori} expectations about the experiment lead the agnostic
experimenter (equal prior expectation), the dogmatic experimenter (one result
is expected with certainty) and the biased experimenter (all intermediate
cases).

Applying this framework to alien hunting is frustrated by three features
that are atypical of a generic scientific experiment. The phenomenon
we seek, alien life, potentially has 1) unbounded explanatory capability
(UEC), 2) unbounded avoidance capacity (UAC) and 3) we, as observers, have
incomplete natural understanding (INU).

We describe a naive experiment that aims to defines its objective as
detecting that which is consistent with alien behaviour, Experiment A,
which is plagued by UEC and UAC. Since effectively any observation of nature
could be deemed to the product of some higher intelligence
\citep{bostrom:2003}, every experiment can be interpreted as a positive
detection and thus A becomes an act of futility. Likewise, a true
null detection is obtainable via UAC, since life could always be present
just somehow evading detection. We conclude that mere \textit{consistency}
with imagined alien behaviour is insufficient as a useful framing of such
experiments, unless that imagination has well-defined limits.

As a refinement, we describe experiment B, where one looks for that which is
inconsistent with lucagenic (originating from life stemming from LUCA) or
abiogenic (originating from processes devoid of life) processes - together
dubbed combigenic. We argue this is plagued by INU and is practically
unrealisable, since one cannot assert with absolute confidence that we are
aware of all possible combigenic phenomena.

Finally, we suggest Experiment C, which seeks that which is inconsistent
\textit{known} combigenic processes and fundamentally shifts the question from
looking for aliens for looking for that which is outside of our current
combigenic understanding. This subtle shift in hypothesis allows for a
quantifiable and well-defined TPP and CPP (assuming the experiment itself is
well understood) and thus a reliable Bayes factor can result. However, it
comes at the cost of avoiding the alien question directly.

We show how this can be re-integrated through Bayesian model selection as a
final step, which has the advantage of clearly elucidating assumptions about
our knowledge of combigenic phenomena. Posing a kind of hierarchical model
selection, one must now attempt to estimate the probability of an
unknown combigenic process generating a phenomenon that would be inconsistent
with known natural processes, $\pdf(\natur|\unnaturX)$. We show how unless
$\pdf(\natur|\unnaturX)<(1+\mathcal{O}_{\mathrm{crit}})^{-1}$ (where
$\mathcal{O}_{\mathrm{crit}}$ is some pre-defined odds ratio threshold
for a ``detection''), a positive detection of aliens is impossible.
Practically speaking, the lesson here is that there's no point looking for
aliens if there is even a small chance that your natural understanding
is incomplete and could confound you.

\subsection{Contextual Discussion}

Put together, we advocate that Experiment C is the most transparent
and realisable framing of the problem. In the other cases, the TPP
and CPP are often unquantifiable since they are predicated upon the
assumed behaviours of alien life, or tacitly (and unrealistically)
depend upon a complete understanding of the Universe.

To close, we briefly discuss interpreting Experiment C in the context
of some possible searches for life. In each case, the primary question
will be - can we assuredly claim that $\pdf(\natur|\unnaturX) \ll 1$?\\

\textbf{Exoplanet biosignatures.} Considerable effort has focussed
on biosignatures which our own Earth presents (e.g. \citealt{sagan:1993}),
especially oxygen and by proxy o-zone \citep{kaltenegger:2010,snellen:2013,
kasting:2014}. In this context, the sub-hypothesis prior becomes
$\pdf(\natur|\unnaturX)\to\pdf(\mathrm{abiotic\,\,O}|\mathrm{O\,\,detected})$.
However, multiple mechanisms for the abiotic production of oxygen have been
identified \citep{worsworth:2014,domagal:2014,lugur:2015,harman:2015},
implying that
$\pdf(\mathrm{abiotic\,\,O}|\mathrm{O\,\,detected})$ could
indeed be too high for oxygen to serve as a viable biosignature. However, in
considering such a claim, clearly the context and properties of the target
planet will significantly influence
$\pdf(\mathrm{abiotic\,\,O}|\mathrm{O\,\,detected})$. Indeed, it was initially
thought that abiotic oxygen could only accumulate to substantial levels outside
of the habitable zone \citep{marais:2002}, but this view has largely
shifted in recent years \citep{schwieterman:2018}. Nevertheless, these mechanisms
should typically be coupled to other indicators besides from the oxygen itself.
For example, one expects to see CO in the case of CO$_2$ photolysis
\citep{harman:2015,schwieterman:2016}. Checking for such contextual information
could thus be used to assess
$\pdf(\mathrm{abiotic\,\,O}|\mathrm{O\,\,detected})$, at least in principle.

The challenge is that clearly in the last decade multiple new pathways have
been identified and thus it is possible, perhaps even probable, that additional
latent mechanisms exist, for which we cannot conduct such context checks since
we haven't even yet predicted their existence! This problem may be somewhat
alleviated by looking for pairs of biosignature molecules, rather than just
one, with the typical framing here to be look for a two molecules in an extreme
thermodynamic disequilibrium \citep{lederberg:1965,lovelock:1965,segura:2005}.
Indeed, oxygen and methane is the most frequently discussed pair to seek
\citep{hitchcock:1967,sagan:1993}. However, even this does not allow us to
confidently assert $\pdf(\natur|\unnaturX)\ll1$ though, since a direct image
of an Earth-like planet is unlikely to resolve close-in moons. This means
that a methane-rich moon (e.g. Titan) orbiting a desiccated, photolysed
planet could produce the pair without any biology involved \citep{rein:2014}
The probability of such a scenario is wholly unknown at this point and
thus we cannot currently confidently assert $\pdf(\natur|\unnaturX)\ll1$,
and thus consequently hope to detect life until this is resolved. Fortunately,
we can conceive of observations to address this specific concern \citep{hek}
but we may always be somewhat in the shadow of the unknown unknowns that
may yet push $\pdf(\natur|\unnaturX)$ perilously high.

\textbf{Unidentified Flying Objects (UFOs).} As a counter-point to the
relatively ``mainstream'' previous example, we consider next the highly
controversial issue of UFOs. Consider an observer who reports seeing an
aerial phenomenon that is not identifiable (which defines a UFO). An
investigation is then established to evaluate the merit of the hypothesis
that this was some kind of manifestation of alien activity. As is generally
true, framing the problem in terms of Experiment A leads to an impotent
experiment and UEC is vivid when consider hyper-advanced aliens. Even if the
observer is assumed to have robotic-like objectivity (never mis-remembers,
deceives, hallucinates, etc), Experiment B is unrealisable since we cannot
legitimately claim they (or subsequent investigators) have complete
knowledge of every balloon, bird, drone, spy satellite, etc in and around
our atmosphere at all times, nor can we claim the observer can never be
deceived by perspective illusions/distortions. So, as is generally true,
we must frame this via Experiment C, which admits the possibility that
UFOs need not be aliens (even if the UFO is indeed real).

Proceeding with Experiment C reveals an immediate problem. Let us generously
assume that the observer represents a perfect experiment ($\TPP=1$ and
$\CPP=0$), such that $\Kpos=\infty$. Although the existence of the UFO
is now established without doubt, this does not yet establish aliens. The case
for aliens comes down assessing $\pdf(\natur|\unnaturX) \to
\pdf(\mathrm{non-alien\,\,phenomenon|unidentified\,\,object})$, which recall
must be close to zero for the alien hypothesis to win out. As currently stated,
it's trivial to dismiss the claim as untenable, since one cannot plausibly
contend that an observer's inability to identify something means
there's no alternative explanation besides from aliens.

To address this, the case could be sharpened by adding the extra constraint
that the UFOs engage in some kind of activity that no known phenomenon can
perform. Now, $\pdf(\natur|\unnaturX)$ can be legitimately argued to be
potentially small, depending on the context and property in question - and
indeed this reasoning motivates the Galileo Project for example
\citep{loeb:2023}. When it comes to eyewitness accounts of such phenomenon,
a basic challenge is that no-one can realistically claim to be a perfect
experiment with $\CPP=0$. Humans will occasionally see misidentify things,
or even mis-remember or imagine them \citep{lynn:1996,clancy:2002}. In
principle, one could imagine determining the $\CPP$ of a group of
observers by placing them in some kind of control environment where have
certainty that aliens are not present and then measuring how often they
make spurious sightings. The problem is that the presumed extreme
capabilities, high visitation frequency and global presence of the
hypothesised aliens means no such control environment exists, unless one
first rejects the alien hypothesis. Going from eyewitness accounts to
recorded images (e.g. Galileo Project; \citealt{loeb:2023}) does not
trivially solve this issue either, since if aliens are ubiquitous in our
skies they represent a background thereby obscuring our ability to measure
the $\CPP$.

It is beyond the scope of this paper to attempt to solve these problems,
nor do we do claim they are unsolvable, but we do claim that if the
$\CPP$ and $\TPP$ cannot be quantified, there is no path forward to
determining $\Kpos$ and thus we can never hope to confirm aliens via
such an experiment (even if $\pdf(\natur|\unnaturX) \to 0$).

\textbf{Narrow-Band Radio Signal:} As a final much briefer case study, we
consider a narrow-band radio signal as a technosignature, one of the first
methods suggested to look for alien life \citep{cocconi:1959,drake:1961}. Very
narrow-band radio signals are not known to be produced naturally, nor is there
any known natural mechanism to generate such a signal. However, legitimate
concerns might still persist about whether such a signal is truly alien.

The first is exemplified by the Wow! signal \citep{kraus:1979} -
which exhibits a narrowband signal ($\lesssim 10$\,kHz) around the
21\,cm line of high signal-to-noise; implying a low $\CPP$ and high
$\TPP$. Despite this, doubts persist about its reality since it
cannot be established whether the signal is genuinely extrasolar
or not from the available data nor has it repeated \citep{gray:2002}.
In principle, this is a solvable problem though, for example
by looking for interstellar scintillation that could verify its
extrasolar origin \citep{brzycki:2023}.

The second concerns $\pdf(\natur|\unnaturX)$ - how confident can we truly be
that some as-yet-undiscovered process generates narrow-band radio signals? A
lucagenic example might be a spy satellite broadcasting on this band, whereas
an abiogenic example might be some previously unknown astrophysical phenomenon.

Of all proposed bio/technosignatures, we would argue that it's here where radio
SETI comes into its own, at least potentially. For the long held dream of radio
SETI is not just a radio ping but a detailed information-rich message, whose
high density and artificial nature could be truly unambiguous. In this case, we
have the rare example where we can perhaps already confidently assert that such
a message would satisfy $\pdf(\natur|\unnaturX) \ll 1$.

\section*{Acknowledgements}
We thank the anonymous reviewer for their constructive comments.
DK thanks M. Sloan, D. Daughaday, A. Jones, E. West, T. Zajonc, C. Wolfred, L. Skov, G. Benson, A. de Vaal, M. Elliott, M. Forbes, S. Lee, Z. Danielson, C. Souter, M. Gillette, T. Jeffcoat, J. Rockett, T. Donkin, A. Schoen, J. Black, R. Ramezankhani, S. Marks, N. Gebben, M. Hedlund, D. Bansal, J. Sturm, Rand Corp., L. Deacon, R. Provost, B. Sigurjonsson, B. P. Walford, N. De Haan, J. Gillmer, E. Garland, A. Leishman, T. Queen Rd. Fnd. Inc, B. T. Pearson, S. Thayer, B. Seeley, F. Blood \& I. Williams.

%






\bibliography{manuscript}{}
\bibliographystyle{aasjournal}



\end{document}